\documentclass[12pt,a4paper]{article}
\usepackage{amsfonts,latexsym}
\usepackage{graphicx,color}
\usepackage{dcolumn}
\usepackage{graphicx}
\usepackage{amsmath}
\usepackage{amsfonts}
\usepackage{amssymb}
\usepackage{psfrag}
\usepackage{wrapfig}
\usepackage{subfigure}
\usepackage{makeidx}

\renewcommand{\thefootnote}{\fnsymbol{footnote}}

\begin{document}

\vspace{12mm}

\begin{center}
{{{\Large {\bf New  scalarization of the Einstein-Euler-Heisenberg black hole}}}}\\[10mm]
{Lina Zhang$^1$\footnote{e-mail address: linazhang@hnit.edu.cn}, 
De-Cheng Zou$^2$\footnote{e-mail address: dczou@jxnu.edu.cn} and Yun Soo Myung$^{3}$\footnote{e-mail address: ysmyung@inje.ac.kr}}\\[8mm]

{${}^1$College of Science, Hunan Institute of Technology, Hengyang 421002, China\\[0pt]}

{${}^2$College of Physics and Communication Electronics, Jiangxi Normal University, Nanchang 330022, China\\[0pt]}

{${}^3${Center for Quantum Spacetime, Sogang University, Seoul 04107, Republic of  Korea\\[0pt] }}

\end{center}
\vspace{2mm}

\begin{abstract}
The spontaneous scalarization of the  Einstein-Euler-Heisenberg (EEH) black hole is performed in the EEH-scalar theory by introducing an exponential scalar coupling (with $\alpha$ coupling constant) to the Maxwell term.
Here, the EEH black hole as a blad black hole is  described by mass $M$ and  magnetic charge $q$  with an action parameter $\mu$. A choice of $\mu=0.3$ gurantees a single horizon with unrestricted magnetic charge $q$. 
The onset scalarization of this black hole appears  for a positive coupling $\alpha$ with an   unlimited magnetic charge $q$. However, there exists a difference  between $q\le1$ and $q>1$ onset scalarizations. 
We notify the presence of  infinite branches labeled by the number of $n=0,1,2,\cdots$  of  scalarized charged  black holes by taking into account the scalar seeds  around  the EEH black hole.
We find that   the $n=0$  fundamental branch of all scalarized  black holes  is stable against the radial perturbations, while the $n=1$ excited branch is unstable.
\end{abstract}

\vspace{1.5cm}

\hspace{11.5cm}
\newpage
\renewcommand{\thefootnote}{\arabic{footnote}}
\setcounter{footnote}{0}

\section{Introduction}

No-hair theorem states that a black hole is completely described by the mass ($M$), electric charge ($Q$), and rotation parameter ($a$)~\cite{Ruffini:1971bza}.
If a scalar field is minimally coupled to  gravitational and electromagnetic  fields, it could not survive  as an equilibrium configuration around the black hole, which describes  no-scalar hair theorem~\cite{Herdeiro:2015waa}.

Introducing  nonminimal couplings, however,  analytic solutions of black hole with scalar hair have been found.
Considering  a conformal scalar coupling to the Ricci scalar ($\phi^2 R/6$),
 the BBMB black hole with secondary scalar hair has been found~\cite{Bocharova:1970skc,Bekenstein:1974sf}, regarding as an extremal black hole 
 (outer horizon=inner horizon).
If one introduces a dilatonic coupling ($e^\phi$) coupling to the Maxwell term, one obtained the well-known GMGHS black hole solution with secondary dilatonic  hair~\cite{Gibbons:1987ps,Garfinkle:1990qj,Gibbons:1982ih,Pope:2024ncb}, but  its inner horizon disappears. These asymptotically flat black hole solytions support the no scalar-haired inner horizon theorem. This theorem implies  that  there exist no inner (Cauchy) horizons for spherically symmetric  black holes with nontrivial scalar (dilaton)  hairs~\cite{Cai:2020wrp,Devecioglu:2021xug,Dias:2021afz}. In other words,  this states a close connection between scalar hair and inner horizon.

Many  black hole solutions with scalar hair were numerically constructed  from the Einstein-Gauss-Bonnet-scalar theory~\cite{Doneva:2017bvd,Silva:2017uqg,Antoniou:2017acq} and Einstein-Maxwell-scalar (EMS) theory~\cite{Herdeiro:2018wub} by introducing  the nonminimal coupling function $f(\phi)$ to the Gauss-Bonnet term  and Maxwell term, respectively~\cite{Myung:2018vug,Myung:2018jvi,Myung:2019oua}. This is called spontaneous scalarization where there is no room to include the inner horizon because they belong to series solutions.
It is worth noting that the onset of spontaneous scalarization  is surely captured by its linearized scalar theory. This  was usually implemented by the tachyonic instability of a linearized scalar  around  the bald [Schwarzschild and Reissner-Nordstr\"om (RN)] black holes without scalar hair.  Nowadays, we are getting into  the evasion era of  the no-hair theorem with the no scalar-haired inner horizon theorem.

In this work, we consider the Einstein-Euler-Heisenberg-scalar (EEHS) theory  to investigate the spontaneous scalarization of its EEH black hole with the Euler-Heisenberg parameter $\mu$.
In  case of $\mu\le 0.08$ with black hole mass $M=1$, it is important to note that  one  could obtain  a single horizon with unlimited magnetic charge ($q>0$), comparing to $q\le 1$ for RN black hole.  Hence, this satisfies the no scalar-haired inner horizon theorem automatically.
At this stage, we note that a negative potential-induced scalarization of EEH black hole without any scalar coupling function $f(\phi)$ led to a single branch of unstable scalarized EEH black holes~\cite{Guo:2025ksj}. 
Introducing an exponential scalar coupling $f(\phi)=e^{-\alpha \phi^2}$  with a scalar coupling parameter $\alpha$ to the Maxwell term $\mathcal{F}$,  the onset scalarization of this black hole occurs for a positive coupling $\alpha$ and an  unlimited magnetic charge $q>0$.
In the sutdy of onset scalarization, we introduce three conditions  for tachyonic instability with $\alpha\ge \alpha_{\rm i}$: 
 sufficient condition $\alpha_{\rm sEEH}(M,q)$, instability condition $\alpha_{\rm in}(M,q)$, and threshold of instability $\alpha_{\rm th }(M,q)$.  Here,  the first and second  are found analytically but they are approximate quantities, while the latter is obtained by numerically and it is a exact quantity. 
From $\alpha_{\rm sEEH}(M,q)$ and $\alpha_{\rm in}(M,q)$, one finds a different behavior of the onset scalarization  between $q\le 1$ and $q>1$. The former is simlar to the onset scalarization of conventional RN black holes, whereas the latter is regarded as   a newly scalarization.   We stress again that the latter is surely considered  as  the new domain of spontaneous scalarization  when choosing an  unlimited magnetic charge $q>0$.
We  check the existence  of  infinite  branches ($n=0,1,2,\cdots$) for  scalarized charged  black holes by studying the scalar seeds around  the EEH black holes. 

We  will obtain   scalarized charged black holes for different $q=0.5,2,20$ in the $n=0$ fundamental branch  by solving full equations.
The stability of the $n=0$  branch  with respect to radial perturbations will be reached by examining the qualitative behavior of the scalar potentials around the $n=0$ branch  as well as by obtaining  exponentially growing (unstable) modes for $s$-mode scalar  perturbation.
We expect to find that the $n=0$ branch is stable, implying that its astrophysical applications might  be put to the test.

The organization of the present work is as follows.
In section 2, we introduce  the EEHS theory and  the EEH black holes.
In section 3,  we study  the  onset scalarization  around the EEH black holes extensively  by emplying   analytic and nemerical methods to handle the linearized scalar theory.
We discuss onset scalarization for electrically charged black holes in section 4.
Section 5 is devoted to exploring scalarized EEH black holes with $q=0.5,2,20$ in the $n=0$ funmadental branch.  The stability  analysis of the  $n=0,1$ branches will be performed by introducing the radial perturbations in section 6.
We obtain scalarized EEH black holes with both scalar couplings in section 7.
Finally, we would like to mention discussions on our results in section 8.

\section{EEH black holes}

 The Einstein-Euler-Heisenberg-scalar (EEHS) action is introduced with an action parameter $\mu$ for a nonlinear electrodynamics term $\mathcal{F}^2$
\begin{equation}
S=\frac{1}{16 \pi}\int d^4 x\sqrt{-g}\Big[ R-2\partial_\mu \phi \partial^\mu \phi-e^{-\alpha \phi^2} \mathcal{F}+\mu \mathcal{F}^2\Big],\label{Act1}
\end{equation}
where $\alpha$ is a scalar coupling constant to the Maxwell term, $\mathcal{F}=F_{\mu\nu} F^{\mu\nu}$.

Before we proceed, we  note that scalarized dyonic  black holes were obtained from the EMS  with an additional  quasi-topological term of $\mathcal{F}^2-2F^{(4)}$ with $F^{(4)}=F^\mu_\nu F^\nu_\rho F^\rho_\sigma F^\sigma_\mu$~\cite{Myung:2020ctt} and scalarized Bardeen  black holes were recently found from the Einstein-scalar coupling to a nonlinear electrodynamics term [$\frac{(q^2\mathcal{F}/2)^{5/4}}{(1+\sqrt{q^2\mathcal{F}/2})^{5/2}}$] which provides a regular black hole as a bald one~\cite{Zhang:2024bfu}. Their $n=0$ branches turned out to be stable against radial perturbations. 

The Einstein  equation can be obtained  from the action Eq.(\ref{Act1})  under  the variation with respect to the metric tensor $g_{\mu\nu}$ as
\begin{eqnarray}
 G_{\mu\nu}=2\Big[\partial _\mu \phi\partial _\nu \phi -\frac{1}{2}(\partial \phi)^2g_{\mu\nu}+T_{\mu\nu}\Big], \label{equa1}
\end{eqnarray}
where the energy-momentum tensor is given by 
\begin{eqnarray}
T_{\mu\nu}&=&e^{-\alpha \phi^2}\Big(F_{\mu\rho}F_{\nu}~^\rho-\frac{1}{4}\mathcal{F}g_{\mu\nu}\Big)
          -2\mu\Big(\mathcal{F}F_{\mu\rho}F^\rho_\nu-\frac{1}{8}\mathcal{F}^2g_{\mu\nu}\Big)\label{emten}.
\end{eqnarray}
The Maxwell equation takes the form 
\begin{eqnarray} \label{M-eq}
&&\nabla_\mu (F^{\mu\nu}-2\mu\mathcal{F}F^{\mu\nu})=2\alpha \phi\nabla_{\mu} (\phi)F^{\mu\nu}.\label{M-eq1}
\end{eqnarray}
On the other hand, the scalar equation involes an effective mass term due to the nonminimal scalar coupling to the Maxwell term as
\begin{equation}
\nabla^2 \phi +\frac{\alpha}{2}\mathcal{F}e^{-\alpha \phi^2}\phi=0 \label{s-equa}.
\end{equation}
Introducing  the mass function $\bar{m}(r)$ together with the gauge field $\bar{A}_{\hat{\varphi}}=-q\cos\theta$ and the hairless scalar $\bar{\phi}=0$,  one finds  the Einstein equation 
\begin{equation}
\bar{m}'(r)=\frac{q^2}{2r^2}-\mu \frac{q^4}{r^6}. \label{mass-eq}
\end{equation}
Its spherically symmetric  solution is given by  the EEH black hole~\cite{Yajima:2000kw,Amaro:2020xro,Breton:2021mju,Allahyari:2019jqz}  as
\begin{eqnarray}
ds^2=\bar{g}_{\mu\nu}dx^\mu dx^\nu=-f(r) dt^2+\frac{dr^2}{f(r)} +r^2d\Omega^2_2,   \label{EEH-s}
\end{eqnarray}
where the metric function takes the form with the Maxwell term 
\begin{equation} \label{metric-func}
 f(r)\equiv1-\frac{2\bar{m}(r)}{r}=1-\frac{2M}{r}+\frac{q^2}{r^2}-\frac{2\mu q^4}{5r^6},\quad \bar{\mathcal{F}}=\frac{2q^2}{r^4}.
\end{equation}
It is meaningful to note that the EEH black hole contains  three  parameters $\{M,q,\mu\}$.
Here,  $M$  represents  the ADM mass, $q$ denotes  the magnetic  charge, and $\mu$ is the action  parameter.   In case of $\mu\le 0.08$ with $M=1$, it was shown that there exist three  horizons with one point at $q=0$~\cite{Myung:2025zxu}, leading to a complicated analysis.
In the present work, it would be better to introduce  a single  horizon  which  satisfies the no scalar-haired inner horizon theorem automatically~\cite{Cai:2020wrp}.
If one  chooses $\mu= 0.3$ appropriately,  one finds  the single horizon.  
In (Right) Fig. 1,  we represent  three metric functions $f(r,M=1,q,\mu=0.3)$ as functions of $r\in[0.5,20]$ with $q=0.5,2,20$.   They cross $r$-axis at $r=1.87(q=0.5), 0.89(q=20),2.63(q=20)$, which represent the event horizons $r_+(M=1,q)$ at $q=0.5,2,20$.

\begin{figure*}[t!]
   \centering
  \includegraphics[width=0.4\textwidth]{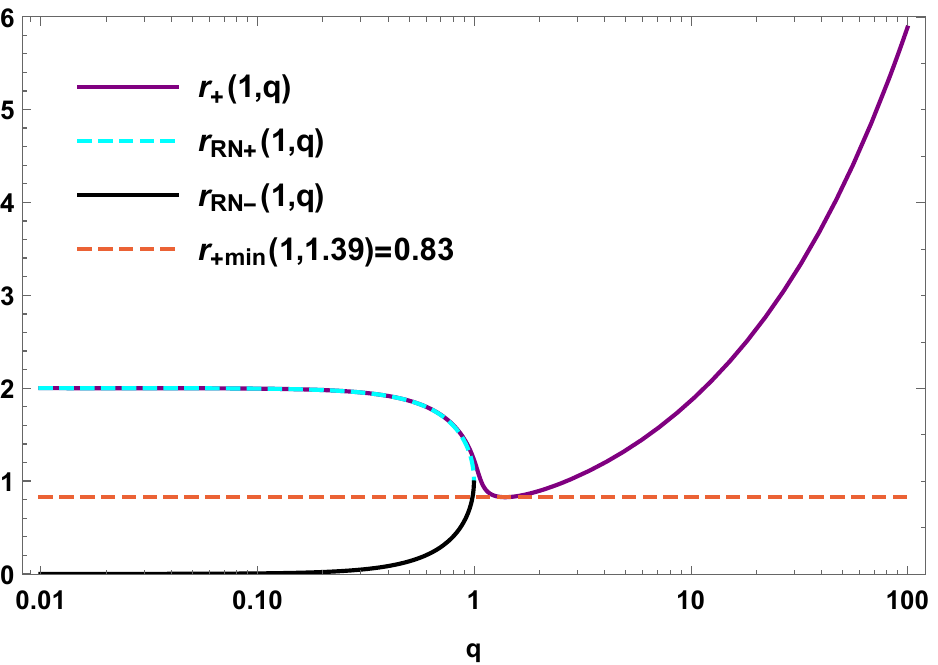}
   \hfill%
\includegraphics[width=0.4\textwidth]{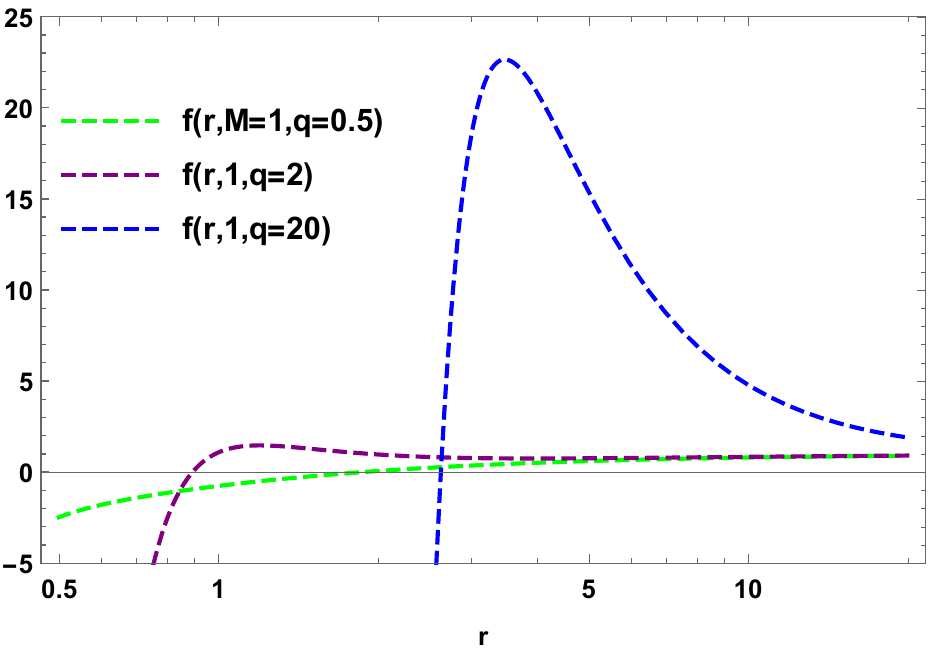}
\caption{(Left) Two outer horizons $r_+(1,q)>r_{\rm RN+}(1,q\in[0,1])$ with an inner horizon $r_{RN-}(1,q\in[0,1])$. $r_+(1,q)$ has the minimum of $r_+(1,1.39)=0.83$ and then, it is an increasing function of $q$. We note $r_+(1,100)=5.88$.
(Right) Metric functions $f(r,M=1,q,\mu=0.3)$ as functions of $r\in[0.5,20]$ with $q=0.5,2,20$.   
They cross $r$-axis at $r=1.87(q=0.5),~ 0.89(q=2),~2.63(q=20)$, representing  three event horizons $r_+(M=1,q)$ at $q=0.5,2,20$. }\label{fig1}
\end{figure*}

As is shown in (Left) Fig. \ref{fig1}, we display the single horizon   $r_{+}(M=1,q)$ to $f(r)=0$.
It is worth noting that  there is no  theoretical constraint on restricting  the magnetic charge $q$ and thus,  $r_{+}(1,q)$ becomes  a continuous  function of $q$: it takes  the minimum of $r_+(1,1.39)=0.83$ and then, it is an increasing function of $q$ with  $r_+(1,q=100)=5.88$.
It differs quitely  from the well known RN black hole ($\mu=0$ case) possessing  outer/inner horizons $r_{RN\pm}(M=1,q)=1\pm\sqrt{1-q^2}$ and having a confined region of $q\in[0,1]$.
We note $r_+(1,q)$ and $r_{RN+}(1,q)$ are similar to each other only for $q\in[0,1]$.

\section{Onset of spontaneous scalarization}

To study the onset of spontaeous scalarization, we need to introduce the linearized scalar theory.
The linearized metric theory was used to perform the stability of EEH black holes~\cite{Luo:2022gdz}.
If one introduces a scalar perturbation around the EEH black hole background as 
\begin{equation}
\phi=0+\delta \varphi,
\end{equation}
its  linearized scalar equation is given by 
\begin{equation}
\Big(\bar{\nabla}^2+\frac{ \alpha q^2}{r^4}\Big)\delta \varphi=0.\label{per-eq}
\end{equation}

\begin{figure*}[t!]
   \centering
  \includegraphics[width=0.32\textwidth]{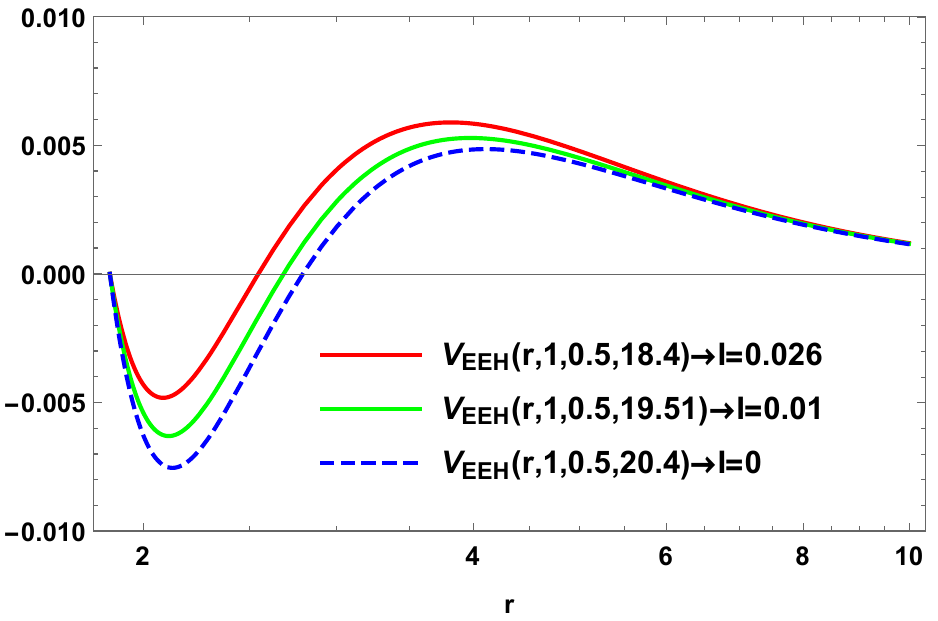}
   \hfill%
\includegraphics[width=0.3\textwidth]{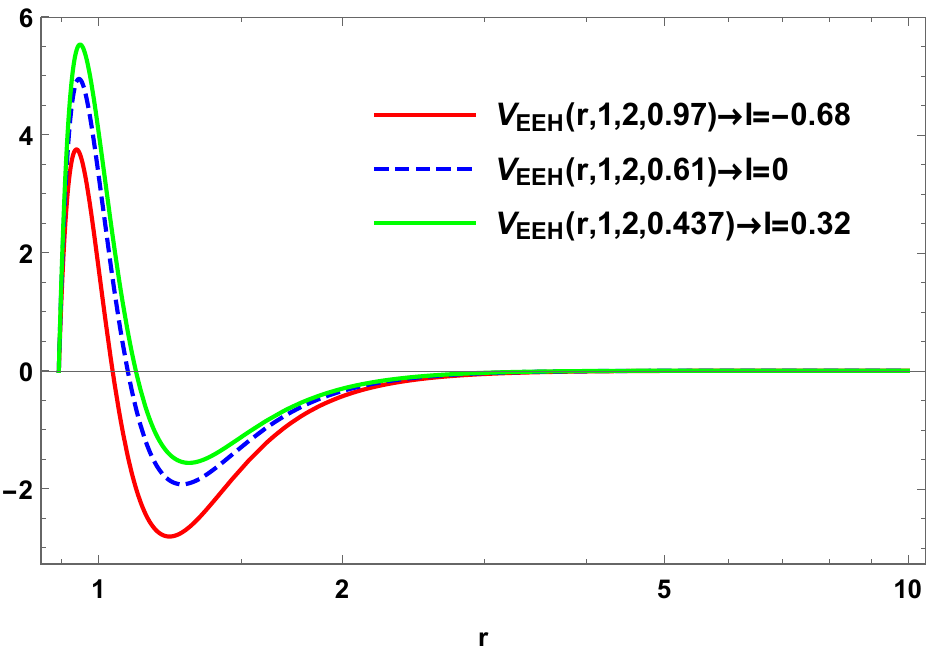}
\hfill%
\includegraphics[width=0.3\textwidth]{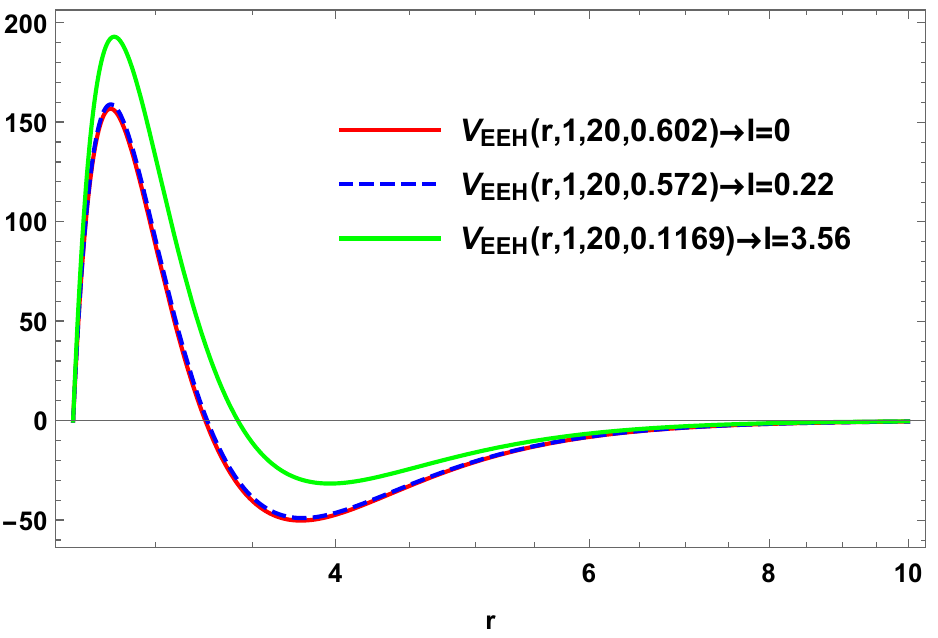}
\caption{Potential $V_{\rm EEH}(r,M=1,q,\alpha)$ and its integration $I(M=1,q,\alpha)$ with $q=0.5,2,20$. (Left) Three $\alpha$-dependent potentials $V_{\rm EEH}(r,M=1,q=0.5,\alpha)$ as functions of
$r\in [r_+(1,0.5)=1.866,10]$  with $\alpha=18.4(=\alpha_{\rm in}),~19.5102(=\alpha_{\rm th}),~20.4(=\alpha_{\rm sEEH})$. (Middle)  Three   potentials
$V_{\rm EEH}(r,1,q=2,\alpha)$ as functions of $r\in [r_+(1,2)=0.893,5]$  with $\alpha=0.437(=\alpha_{\rm th}),~0.61(=\alpha_{\rm sEEH}),~0.97(=\alpha_{\rm in})$. (Right) Three   potentials
$V_{\rm EEH}(r,1,q=20,\alpha)$ as functions of $r\in [r_+(1,20)=2.63,10]$  with $\alpha=~0.1169(=\alpha_{\rm th}),~~0.572(=\alpha_{\rm in}),~0.602(=\alpha_{\rm sEEH})$. If $q>1.14,17.5$, their roles of $\alpha_{\rm in}$, $\alpha_{\rm th}$,  and $\alpha_{\rm sEEH}$ are exchanged.  }\label{fig2}
\end{figure*}
For our purpose, we introduce  the separation of variables
 \begin{equation}
 \delta \varphi(t,r,\theta,\hat{\phi}) = \varphi(r) Y_{lm}(\theta) e^{i m\hat{ \phi}} e^{-i\omega t},\quad \varphi(r)=\frac{u(r)}{r}.
\end{equation}
In coorperation with   the tortoise coordinate $r_*=\int \frac{dr}{f(r)}$, Eq. (\ref{per-eq}) is converted into  the Schr\"odinger equation for $u(r)$ 
\begin{equation}
\frac{d^2 u(r)}{dr^2_*} +\Big[\omega^2-V_{\rm EEH}(r,M,q,\alpha)\Big] u(r) =0, \label{sch-eq}
\end{equation}
where  the $s(l=0)$-mode potential is given by
\begin{eqnarray}
V_{\rm EEH}(r,M,q,\alpha)&=&f(r)\Big[\frac{2M}{r^3}-\frac{(2+\alpha)q^2}{r^4}+\frac{3.6 q^4}{5r^8}\Big] \to f(r) \cdot  v_{\rm EEH}(r).\label{EEH-P}
\end{eqnarray}
See  Fig. \ref{fig2}, for its $\alpha$-dependent  potential forms $V_{\rm EEH}(r,M=1,q,\alpha)$ with $q=0.5,2,20$.

We wish to compute three quantities of coupling constant $\alpha$: sufficient condition $\alpha_{\rm sEEH}(1,q)$, instability condition $\alpha_{\rm in}(1,q)$, and threshold of instability  $\alpha_{\rm th}(1,q)$, where the first two are approximate results obtained from the scalar potential analytically, while the last is the exact value determined  either by solving Eq.(\ref{sch-eq}) with $\omega=i\Omega$  or by solving the static linearized equation numerically.

First of all, it is easy to compute  the sufficient condition for tachyonic instability given by~\cite{Dotti:2004sh}
\begin{equation}
\int_{r_+(M,q)}^\infty \Big[\frac{V_{\rm EEH}(r)}{f(r)}\Big]dr\to \int_{r_+(M,q)}^\infty v_{\rm EEH}(r)dr\equiv I<0
\end{equation}
which determines the upper bound of $\alpha_{\rm sEEH}(M,q)$ in the instability condition for $q<1.14$
\begin{equation}
\alpha_{\rm in}(M,q)<\alpha_{\rm th}(M,q)<\alpha_{\rm sEEH}(M,q). \label{ineq-R}
\end{equation}
This sequency of instability is usually suitable for the onset scalarization of RN black holes found in the EMS theory~\cite{Myung:2018vug}. 
On the other hand, one finds  other sequency of the inequality  for $1.14<q<17.5$
\begin{equation}
\alpha_{\rm th}(M,q)<\alpha_{\rm sEEH}(M,q)<\alpha_{\rm in}(M,q). \label{ineq-M}
\end{equation}
In addition, for $q>17.5$, one obtains  another sequency of the inequality as
\begin{equation}
\alpha_{\rm th}(M,q)<\alpha_{\rm in}(M,q)<\alpha_{\rm sEEH}(M,q). \label{ineq-L}
\end{equation}
It is worth noting  that Eqs.(\ref{ineq-M}) and (\ref{ineq-L}) are new and appear for the onset of spontaneous scalarization in the EEH black holes with unlimited magentic charge $q$.
(Left) Fig. \ref{fig2} for $q=0.5$ implies  that  the threshold of instability $\alpha_{\rm th}(1,0.5)=19.5102$ is   between $\alpha_{\rm in}=18.4$ and  $\alpha_{\rm sEEH}=20.4$.
Differently, we observe $\alpha$-dependent potentials for $q=2,20$ as shown in (Middle/Right) Fig. \ref{fig2}. Also, we observe that``$-+$" regions  appear in the near-horizon  for $q=0.5$, while ``$+-$" regions are shown in the near-horizon  for $q=2,15$.

\begin{figure*}[t!]
   \centering
  \includegraphics[width=0.4\textwidth]{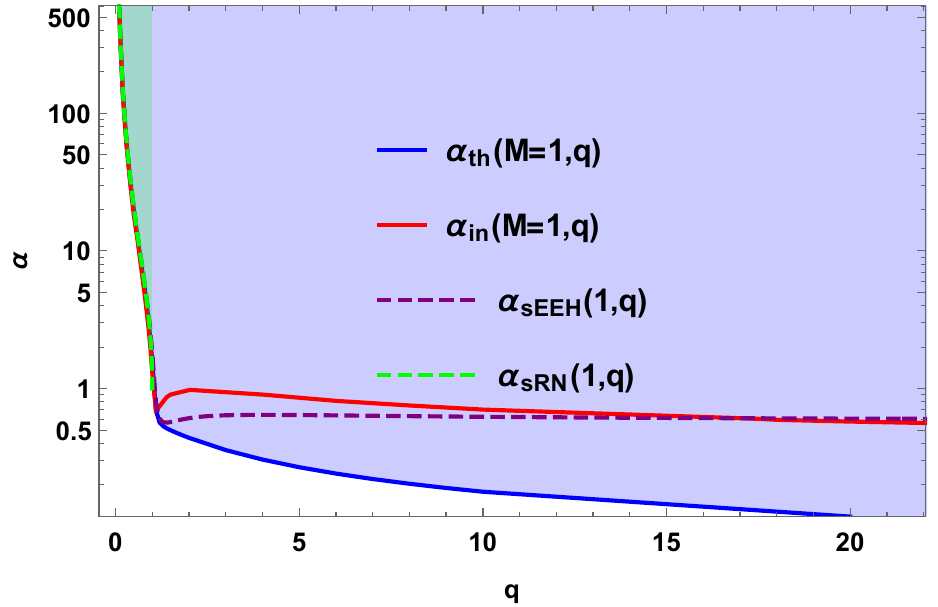}
   \hfill%
\includegraphics[width=0.4\textwidth]{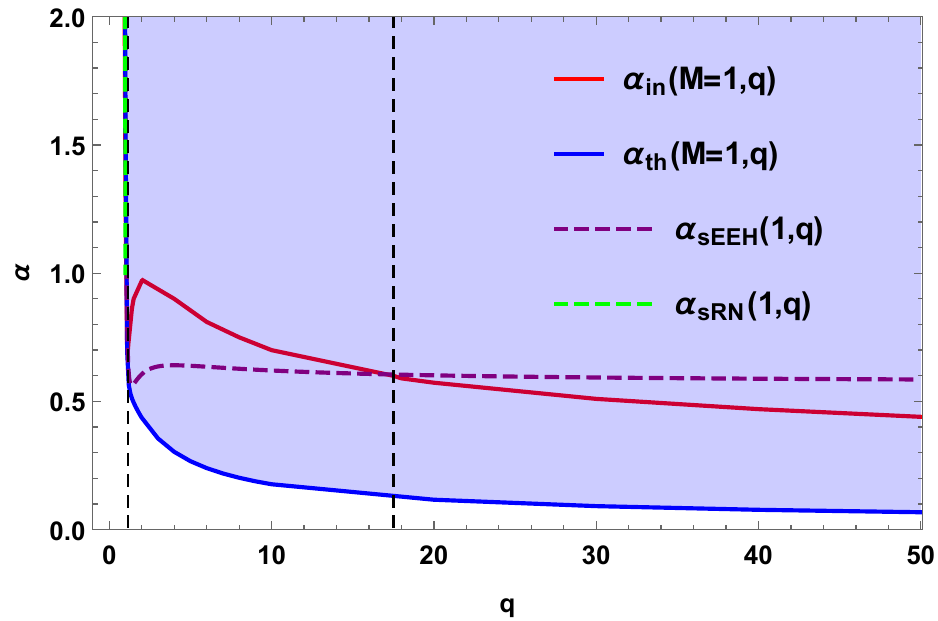}
\caption{(Left) Sufficient condition for tachyonic instability $\alpha_{\rm sEEH}(1,q)$, threshold of instability $\alpha_{\rm th}(1,q)$, and  instability condition $\alpha_{\rm in}(1,q)$ as functions of $q\in[0,22]$ with  $\alpha_{\rm sRN}(1,q\in[0,1])$. It implies a conventional inequality of $\alpha_{\rm in} \le  \alpha_{\rm th}\le \alpha_{\rm sEEH}$ for the green shaded region ($q\in[0,1]$).
 The whole shaded region represents unstable region of $\alpha(1,q)\ge \alpha_{\rm th}(1,q)$. (Right) Their enlarged picture for $\alpha\in[0,2]$.  One finds positive  regions for $\alpha$  with two dashed lines at $q=1.14,17.5$ (crossing points), $\alpha_{\rm sEEH}(1,1000)=0.58$, and $\alpha_{\rm in}(1,1000)=0.22$. One expects to have  other inequality of $\alpha_{\rm th}\le\alpha_{\rm  sEEH}\le\alpha_{\rm in} $ for $1.14<q<17.5$, while  $\alpha_{\rm th}\le\alpha_{\rm  in}\le\alpha_{\rm sEEH} $ for $q>17.5$.  }\label{fig3}
\end{figure*}
At this stage, we present  explicit forms $\alpha_{\rm sEEH}(M=1,q)$ and $\alpha_{\rm sRN}(M=1,q)$~\cite{Myung:2018vug} obtained from the sufficient condition of $I=0$ as
\begin{eqnarray}
\alpha_{\rm sEEH}(1,q\in[0,\infty])&=&-2+\frac{3r_+(1,q)}{q^2}+\frac{0.31q^2}{r^4_+(1,q)}, \label{alp-1} \\
 \alpha_{\rm sRN}(1,q\in[0,1])&=&-2+\frac{3r_{\rm RN+}(1,q)}{q^2}, \label{alp-2}
\end{eqnarray}
which are depicted in (Left) Fig. \ref{fig3}.
 It is important to realize that $\alpha_{\rm sEEH}(1,q)$ is a decreasing function of $q$, but it crosses $\alpha_{\rm in}$ at $q=1.14, 17.5$  which is a new feature for the onset of spontaneous scalarization [see (Right) Fig. \ref{fig3}]. Hence, we choose $q=0.5,2,20$ for real computations. 
Importantly, we observe that for $\alpha\le 1$, the predictions of $\alpha_{\rm sEEH}(1,q)$ and $~\alpha_{\rm in}(1,q)$ are quite diffrent from that for threshold of intability $\alpha_{\rm th}(1,q)$, which implies that two formers may not be suitable for studying the instability of EEH black holes.  However, for  $\alpha>1$, three of $\alpha_{\rm sEEH}(1,q),~\alpha_{\rm in}(1,q),~
\alpha_{\rm th}(1,q)$ with $\alpha_{\rm sRN}(1,q)$ predict the nearly same behavior.


To obtain the instability condition $\alpha_{\rm in}(M,q)$,  the spatially regular  scalar configurations (scalar clouds)  can be obtained  from   Eq.(\ref{sch-eq}) with $\omega=0$ by adopting  the WKB technique~\cite{Hod:2019ulh}.
A second-order WKB method  could be applied for obtaining the bound states of the  potential $V_{\rm EEH}(r_*)$ approximately to yield the quantization condition
\begin{equation}
\int^{r_*^{\rm out}}_{r_*^{\rm in}}dr_* \sqrt{-V_{\rm EEH}(r_*)}=\Big(n-\frac{1}{4}\Big)\pi,\quad n=1,2,3,\cdots, \label{wkb1}
\end{equation}
where $r_*^{\rm out}$ and $r_*^{\rm in}$ are the radial turning points satisfying $V_{\rm EEH}(r_*^{\rm out})=V_{\rm EEH}(r_*^{\rm in})=0$.
We may  express  Eq.(\ref{wkb1}) in terms of the radial coordinate  $r$ as
\begin{equation}
\int^{r_{\rm out}}_{r_{\rm in}}dr \frac{\sqrt{-V_{\rm EEH}(r)}}{f(r)}=\Big(n-\frac{1}{4}\Big)\pi,\quad n=1,2,3,\cdots. \label{wkb2}
\end{equation}
Here, radial turning points $(r_{\rm out},r_{\rm in})$ are determined by the two conditions
\begin{equation}
f(r_{\rm in})=0,\quad \frac{2M}{r^3_{\rm out}}-\frac{(\alpha+2)q^2}{r^4_{\rm out}}+\frac{3.6q^4}{5r^8_{\rm out}}=0,
\end{equation}
which imply
\begin{equation}
r_{\rm in}=r_+(M,q),~\quad r_{\rm out}\simeq \frac{(\alpha+2) q^2}{2M}.
\end{equation}
For large $\alpha(r_{\rm out})$, the WKB integral (\ref{wkb2}) is  approximated by considering the last term in (\ref{EEH-P}) as
\begin{equation}
\sqrt{\alpha}\cdot  q\int^{\infty}_{r_+} \frac{dr}{r^2\sqrt{f(r)}}\equiv\sqrt{\alpha} I_n(M,q) =\Big(n+\frac{3}{4}\Big)\pi,\quad  n=0,1,2,\cdots,
\end{equation}
which could be integrated numerically to yield
\begin{equation} \label{alphan}
\alpha_{\rm in,n}(M,q)=\Big[\frac{\pi(n+3/4)}{I_n(M,q)}\Big]^2,\quad  n=0,1,2,\cdots.
\end{equation}
We plot $\alpha_{\rm in}(M=1,q)[\equiv\alpha_{\rm in,n=0}(1,q)]$ in Fig. \ref{fig3}, which is the lower bound in Eq.(\ref{ineq-R}) for $q<1.14$  as well as the upper bound in Eq.(\ref{ineq-L}) for $1.14<q<17.5$.
It is noted  that $\alpha_{\rm in}(1,q) [\simeq \alpha_{\rm sEEH}(1,q)]$ is a decreasing function of $q$ and  others of $\alpha_{\rm in,n\not=0}(1,q) $ are used to estimate  branch points.
For example, one has $\alpha_{\rm in,0}(1,0.5)=18.4$,  $\alpha_{\rm in,1}(1,0.5)=100 $, and  $\alpha_{\rm in,2}(1,0.5)=247 $, which are similar to exact branch points of $\alpha_0=19.5102$,  $\alpha_1=101.4$, and $\alpha_2=248.73$ for $q=0.5$ obtained from the static linearized equation.

To determine the threshold of tachyonic instability $\alpha_{\rm th}(M,q)$ lastly, we have to solve the second-order differential equation (\ref{sch-eq}).
This   allows an exponentially growing mode of  $e^{\Omega t}(\omega_i=\Omega>0) $ as  an unstable mode for $\omega=\omega_r+i\omega_i$ with $\omega_r=0$.
Here, we choose two boundary conditions: a normalizable
solution of $u(\infty)\sim e^{-\Omega r_*}$  at infinity  and
a power solution of $u(r_+)\sim \left(r-r_+\right)^{\Omega r_+}$  near the outer horizon.
\begin{figure*}[t!]
   \centering
  \includegraphics[width=0.4\textwidth]{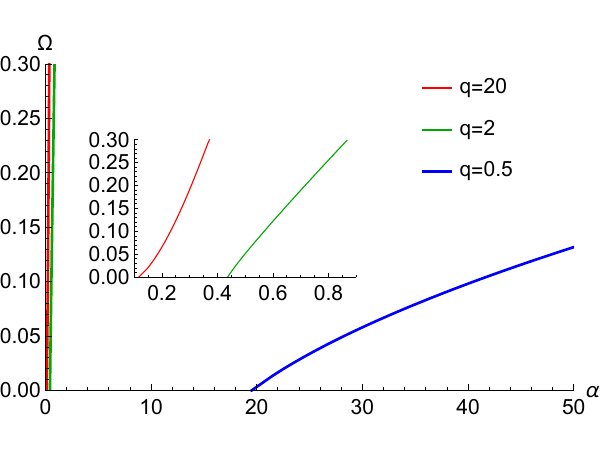}
  \hfill%
\includegraphics[width=0.4\textwidth]{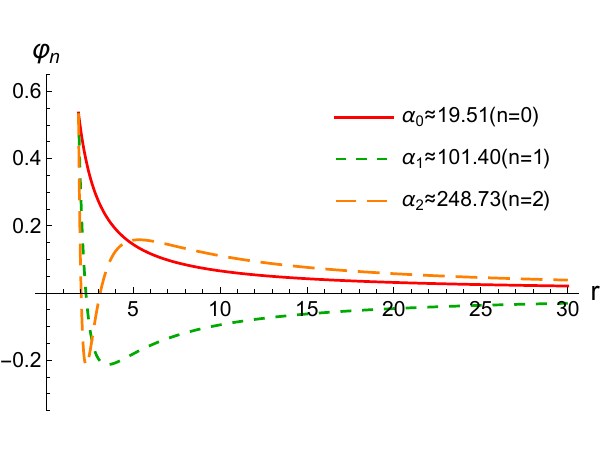}
\caption{
(Left) Three curves of $\Omega$ in $e^{\Omega t}$ as a function of $\alpha$ are used to determine the thresholds of tachyonic instability $[\alpha_{\rm th}(1,q)]$ around the EEH black holes.
We find that  $\alpha_{\rm th}(1,q) =19.5102(q=0.5),~ 0.437 (2),~ 0.1169(20)$ when three curves cross $\alpha$-axis.  
 (Right)   $\varphi(r)=u(r)/r$ as a function of $r\in[r_+=1.87,30]$ for representing  the first three  scalar seeds with $q=0.5$.
$\varphi_n(r)$ is classified by the order number $n=0,1,2$ which is also identified by the number of nodes (zero crossings).  }\label{fig4}
\end{figure*}
We find from (Left) Fig. 4 that  the threshold ($\Omega=0$) of instability  can be determined  as  $\alpha_{\rm th}(M=1,q) = 19.5102(q=0.5),~ 0.437 (2),~ 0.1169(20)$.
Similarly, one obtains $\alpha_{\rm th}(1,q)$ from $q=0.1$ to $q=50$  as shown in  Fig. \ref{fig3}.
This implies that  the EEH black hole is unstable for $\alpha(1,q)\ge\alpha_{\rm th}(1,q)$ precisely, while it is stable for  $\alpha(1,q)<\alpha_{\rm th}(1,q)$.  It is worth mentioning that two of $\alpha_{\rm sEEH}(1,q)$ and $\alpha_{\rm in}(1,q)$ obtained analytically are regarded as  approximate results.

In addition, the other way of getting  $\alpha_{\rm th}(1,q)$  is to solve the  static linearized  equation directly. Also, this can be used to find  the scalar seeds  [$\varphi_n(r)]$ for $n=0,~ 1,~ 2,\cdots$ branches of scalarized EEH black holes.
We consider the  static linearized equation for $\varphi(r)$  around the EEH black hole background
\begin{equation} \label{ssclar-eq}
\frac{1}{r^2}\Big(r^2f(r)\varphi'(r)\Big)'-\Big(\frac{l(l+1)}{r^2}-\frac{\alpha q^2}{r^4}\Big) \varphi(r)=0,
\end{equation}
which describes an eigenvalue problem.  For $l=0$, requiring an asymptotically vanishing scalar [$\varphi(r\to \infty)=0$] imply  that  a smooth scalar
selects  a discrete set of $n=0$, 1, 2, $\cdots$.   Also, it  determines the bifurcation points [$\alpha_n(1,q)$] precisely. First of all, it is confirmed  that  $\alpha_{\rm th}(1,0.5)=\alpha_{0}(1,0.5)$.
We plot scalar seeds $\varphi_n(r)$ with $q=0.5$ as a function of $r$ for  three branches  of  $n=0(\alpha_0=19.5102),~n=1(\alpha_1=101.40),~n=2(\alpha_2=248.73)$ [see  (Left) Fig. \ref{fig4}].
It is important  to note    that the  scalar seed $\varphi_0(r)$ without zero crossing will develop the scalar hair $\phi_0(r)$ which describes the $n=0$  fundamental branch of scalarized charged black holes for  $\alpha\ge \alpha_0(=\alpha_{\rm th})$.
On the other hand,  two scalar seeds  $\varphi_1(r)$ and $\varphi_2(r)$ with zero crossings will develop the scalar hairs $\phi_1(r)$ and $\phi_2(r)$ which describe $n=1$ and $n=2$ excited branches of scalarized charged black holes existing for  $\alpha\ge \alpha_1$ and $\alpha\ge\alpha_2$.
In general,  infinite branches of  $n=0,~1,~2,\cdots$  can be  constructed  from  infinite $\alpha$-bounds of $\alpha\ge \alpha_0,~\alpha\ge \alpha_1$, $\alpha\ge \alpha_2,~\cdots$, respectively.
This describes  briefly  a prescription for obtaining infinite branches of scalarized EEH black holes.
\begin{figure*}[t!]
   \centering
  \includegraphics[width=0.4\textwidth]{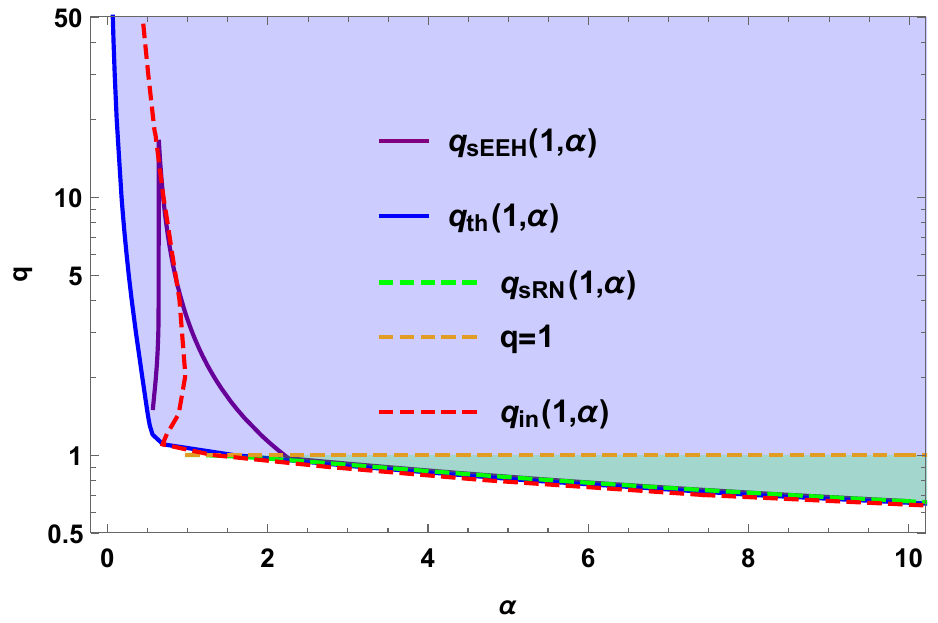}
   \hfill%
\includegraphics[width=0.4\textwidth]{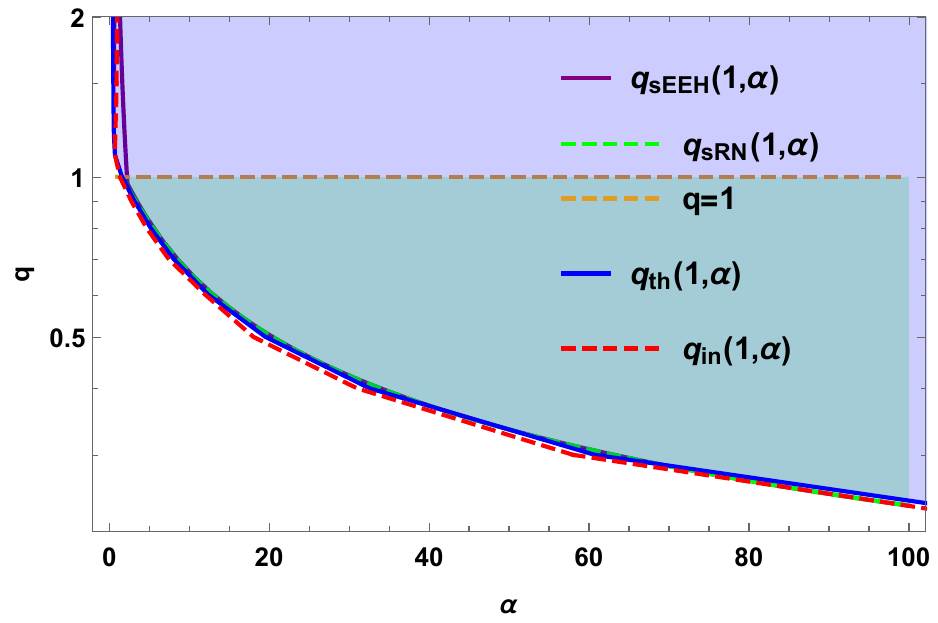}
\caption{(Left) Existence curves $q_{\rm sEEH}(1,\alpha),~q_{\rm th}(1,\alpha)$, and $q_{\rm in}(1,\alpha)$.
The green shaded region [$q_{\rm sRN}(1,\alpha)\le q(1,\alpha)\le 1$] denotes a convential existence region for RN black hole. The whole shaded region represents existence region of $q(1,\alpha)\ge q_{\rm th}(1,\alpha)$, implying  unlimited magenetic charge $q$. (Right) Their enlarged picture for $\alpha\in[0,100]$ vs  $q\in[0,2]$.  }\label{fig5}
\end{figure*}

Finally, we wish to display the existence region for scalarized EEH black holes based on $q_{\rm th}(M=1,\alpha)$ in Fig. \ref{fig5} . Here,  $q_{\rm sRN}(1,\alpha>1)=\frac{\sqrt{3+6\alpha}}{\alpha+2}$ is  found from the condition of $I_{\rm RN}=0$ for $q$, while   $q_{\rm sEEH}(1,\alpha\ge 0.57)$ is obtained from the condition of $I=0$ for $q$, it takes the maximum (=15.2) at $\alpha=0.65$ and then, it is a monotonically decreasing function of $\alpha$. $q_{\rm th} (1,\alpha)$ and $q_{\rm in}(1,\alpha)$ are the inverse functions of $\alpha_{\rm th}(1,q)$ and  $\alpha_{\rm in}(1,q)$, respectively. 
For $q>1$,  $q_{\rm th}(1,\alpha),~q_{\rm sEEH}(1,\alpha),$ and $q_{\rm in}(1,\alpha)$ indicate different predictions. Explicitly,  the predictions of $q_{\rm sEEH}(1,\alpha)$ and $q_{\rm in}(1,\alpha)$ are quite diffrent from that for threshold of intability $q_{\rm th}(1,\alpha)$, implying  that two formers are not appropriate for determing the existence region for scalarized EEH black holes. 

For $q\le1$, however, three of $q_{\rm th}(1,\alpha),~q_{\rm sEEH}(1,\alpha),$ and $q_{\rm in}(1,\alpha)$ with $q_{\rm RN}(1,\alpha)$ predict the nearly same behavior. 
The whole shaded region corresponds to the existence (unstable) region for scalarized EEH black holes without the limitation on the magnetic charge $q$. It includes  a convential existence (unstable) region for RN black hole [green shaded region between $q_{\rm sRN}(1,\alpha)$ and 1].

\section{Onset scalarization of electrically charged black holes }

In this section, we wish to discuss how the onset scalarization of electrically charged black holes works.
For this purpose, we start with the Einstein-Euler-Heisenbegr action~\cite{Magos:2020ykt}
\begin{eqnarray}
S_{\rm EEH}&\equiv&\frac{1}{16 \pi}\int d^4 x\sqrt{-g}(R-\mathcal{L}) \nonumber \\
&=&\frac{1}{16 \pi}\int d^4 x\sqrt{-g}\Big[ R+4F-2a\Big(F^2+\frac{7}{4}G^2\Big)\Big]\label{Act1}
\end{eqnarray}
with $F=F_{\mu\nu}F^{\mu\nu}/4$, $G={}^*F^{\mu\nu}F_{\mu\nu}/4$, and $a=8\mu$.
We use  the $P$-framework with the tensor $P_{\mu\nu}$ to derive an elctrically charged black hole solution
\begin{equation} 
P_{\mu\nu}=-(\mathcal{L}_FF_{\mu\nu}+{}^*F_{\mu\nu}\mathcal{L}_G)=(1-a F)F_{\mu\nu} - \frac{7aG}{4} {^*F}_{\mu\nu},
\label{Pmunu}
\end{equation}
where the subscript $X$ in $\mathcal{L}$ denotes its derivative with respect to $X$. 
We note that two frameworks  $F$ and $P$  correspond to the  Lagrangian and Hamiltonian treatments, respectively.
The two invariants $P$ and $O$ associated to the $P$-framework are defined as
\begin{equation}
P=-\frac{1}{4} P_{\mu\nu} P^{\mu\nu}, \quad 
O= -\frac{1}{4} P_{\mu\nu} {^*P^{\mu\nu}},
\label{Invst}
\end{equation}
where $^*P_{\mu\nu}=\frac{1}{2\sqrt{-g}}\epsilon_{\mu\nu\rho\sigma}P^{\sigma\rho}$ denotes the dual tensor to $P_{\mu \nu}$.
The Legendre transformation of $\mathcal{L}$ is introduced to define the Hamiltonian $\mathcal{H}$
\begin{equation}
\mathcal{H} (P,Q)= -\frac{1}{2}P^{\mu\nu} F_{\mu\nu}-\mathcal{L}.
\end{equation}
Neglecting  higher order terms in $a=8\mu$, the Hamiltonian is given by
\begin{equation} 
\mathcal{H}(P,O)= P- 4\mu \Big( P^2 + \frac{7\mu}{4} Q^2 \Big). \label{Hamiltonian}
\end{equation}
In this case, the Mawell  and Einstein equations take the forms
\begin{eqnarray}
\nabla_\mu P^{\mu\nu}&=&0, \label{motion1} \\
G_{\mu\nu} &=&2 T^e_{\mu\nu},  \label{motion2}
\end{eqnarray}
where the energy momentum tensor $T^e_{\mu\nu}$  in the $P$ framework is given by
\begin{equation}
T^e_{\mu\nu}=\Big[(P_\mu^\beta
P_{\nu\beta}+ Pg_{\mu\nu}) - 8\mu \Big\{ PP_\mu^\beta
P_{\nu\beta}+g_{\mu\nu}\Big(
\frac{3}{2} P^2 + \frac{7}{8}O^2  \Big)\Big\}\Big].  \label{em-ten}
\end{equation}
The electromagnetic field is given by the anti-symmetric tensor $P_{\mu \nu}$ 
\begin{equation}
P_{\mu\nu}= \frac{e}{2r^2}(\delta^1_{\mu} \delta^0_{\nu}- \delta^0_{\mu} \delta^1_{\nu} ), \label{Pmunu01}
\end{equation}
which implies that $P$ and $O$  are determined to be 
\begin{equation}
P=\frac{e^2}{2r^4},\quad  O=0  \label{st} . 
\end{equation}  
Then, the (0,0)-componet of Eq.(\ref{motion2})  leads to 
\begin{equation}
    m'(r)=\frac{e^2}{2r^2}-\frac{\mu e^4}{r^6}
\end{equation}
which implies that 
the metric function for the electric case is given by
\begin{equation}
f_e(r)= 1-\frac{2M}{r}+\frac{e^2}{r^2}-\frac{2\mu e^4}{5 r^6}. \label{gtt}
\end{equation}
This  is the same form of Eq.(\ref{metric-func}) when replacing $e$ by $q$.
To implement onset of scalarization for electrically charged black holes, we may couple $e^{-4\alpha \phi^2}$ to $P$-term  and include $-2(\partial\phi)^2$ by observing the first term in Eq.(\ref{em-ten}).  
In this case, the scalar equation takes the form
\begin{equation}
    \nabla^2 \phi+ 2\alpha P e^{-4\alpha \phi^2} \phi=0.
\end{equation}
This implies that the linearized scalar equation leads to 
\begin{equation}
\Big(\bar{\nabla}^2+\frac{ \alpha e^2}{r^4}\Big)\delta \varphi=0,\label{per-eqe}
\end{equation}
which becomes the same equation (\ref{per-eq}) when replacing $e$ by $ q$.
Hence, we expect to find the same onset of scalarization for magenetically charged black hole which appeared in the previous section.

\section{Scalarized EEH black holes}

 All scalarized EEH black holes can  be generated from the onset of scalarization 
 $\{\varphi_n(r)\}$ in the unstable region of EEH black holes [$\alpha(1,q)\ge \alpha_{\rm th}(1,q)$]. It is clear that the scalarized black hole solutions arise dynamically from the evolution of EEH black hole when scalar cloud $\varphi_n(r)$ plays the role of a seed for $n$-branch.

We wish to obtain  scalarized EEH black holes through spontaneous scalarization by solving full equations.
To this direction,  one considers  the metric and field ansatzes  as~\cite{Herdeiro:2018wub}
\begin{eqnarray}\label{nansatz}
ds^2_{\rm sEEH}&=&-N(r)e^{-2\delta(r)}dt^2+\frac{dr^2}{N(r)}+r^2(d\theta^2+\sin^2\theta d\hat{\varphi}^2) \nonumber \\
N(r)&=&1-\frac{2m(r)}{r},\quad \phi=\phi(r),\quad A= A_{\hat{\varphi}} d\hat{\varphi}.
\end{eqnarray}
Plugging the gauge field ansatz into Eq.(\ref{M-eq}), one finds  a magnetic  potential $A_{\hat{\varphi}}=-q \cos \theta$, which implies  $F_{\theta \hat{\varphi}}=q\sin\theta$ and $\mathcal{F}=2q^2/r^4$.
This means  that it is unnecessary to consider  an approximate solution for $A_{\hat{\varphi}}$. 

Substituting (\ref{nansatz}) into Eqs.(\ref{equa1}) and (\ref{s-equa}) leads to  three coupled equations for $m(r),~\delta(r),~\phi(r)$ as
\begin{eqnarray}
&&q^2e^{-\alpha\phi^2(r)}-\frac{2\mu  q^4}{r^4}-2r^2m'(r)+r^3\left(r-2m(r)\right)\phi'^2(r)=0, \label{neom1}\\
&&\delta'(r)+r\phi'^2(r)=0, \label{neom2}\\
&&\frac{\alpha q^2\phi(r)e^{-\alpha\phi^2(r)}}{r^2}-2[m(r)+rm'(r)-r]\phi'(r) \nonumber \\
&&\quad\quad-r(r-2m(r))[\delta'(r)\phi'(r)-\phi''(r)]=0. \label{neom3}
\end{eqnarray}
It is checked  that Eq.(\ref{neom1}) reduces to Eq.(\ref{mass-eq}) for $\phi(r)=0$.
\begin{figure*}[t!]
   \centering
  \includegraphics[width=0.3\textwidth]{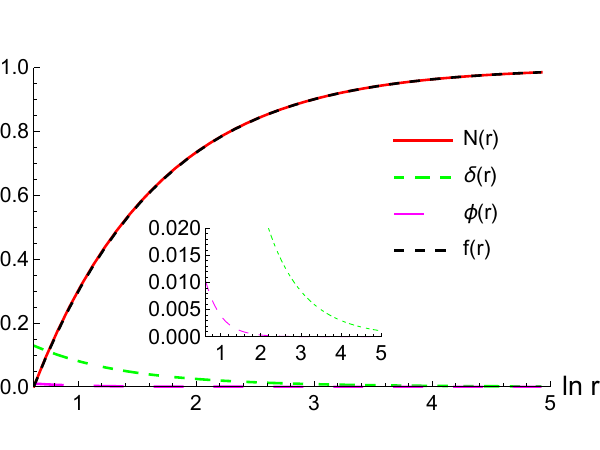}
  \hfill%
\includegraphics[width=0.3\textwidth]{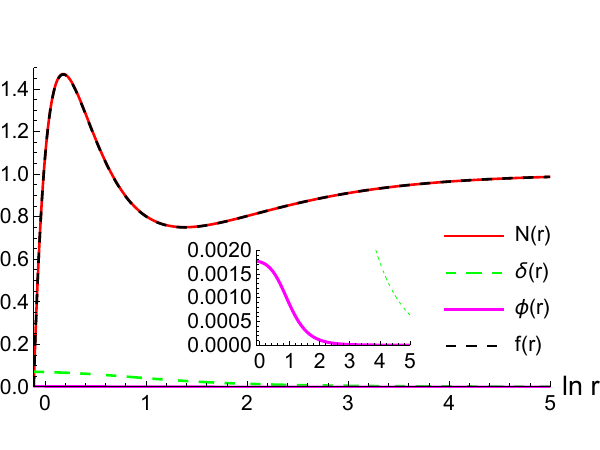}
\hfill%
\includegraphics[width=0.3\textwidth]{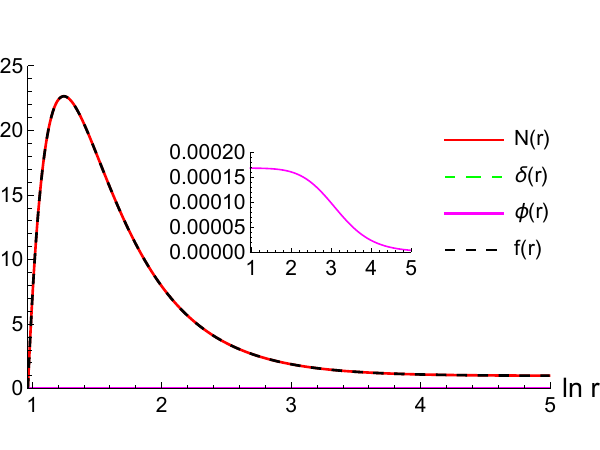}
\caption{Graph  of scalarized EEH black hole  solutions. It shows metric functions $N(r)$, $\delta(r)$, and scalar hair $\phi(r)$.  Here, $f(r)$ represents metric function for EEH black holes as was shown in (Right) Fig. 1.
(Left)  $q = 0.5$,  and $\ln r_+ = 0.624$ for $\alpha = 25$  in the $n = 0$ branch of $\alpha \geq 19.5102$. (Middle) $q = 2$,  and $\ln r_+ = -0.1125$  for $\alpha = 0.438$  in the $n = 0$ branch of $\alpha \geq 0.437$. (Right)  $q = 20$,  and $\ln r_+ = 0.967$ for $\alpha = 0.117$  in the $n = 0$ branch of $\alpha \geq 0.1169$.
 }
\end{figure*}
Assuming  the existence of a single horizon located at $r=r_+$,   an
approximate solution to Eqs.(\ref{neom1})-(\ref{neom3})  takes the form   in the near horizon as
\begin{eqnarray}
m(r)&=&\frac{r_+}{2}+m_1(r-r_+)+\cdots,\label{aps-1} \\
\delta(r)&=&\delta_0+\delta_1(r-r_+)+\cdots,\label{aps-2}\\
\phi(r)&=&\phi_0+\phi_1(r-r_+)+\cdots.\label{aps-3}
\end{eqnarray}
Here, the three coefficients are determined  by
\begin{eqnarray}\label{ncoef}
&&m_1=\frac{e^{-\alpha\phi_0^2}q^2}{2r_+^2}-\frac{\mu q^4}{r_+^6},\quad \delta_1=-r_+\phi_1^2,\quad \phi_1=\frac{\alpha \phi_0 q^2}{r_+[q^2-e^{\alpha\phi_0^2}(2\mu q^4/r_+^4+r^2_+)]}.
\end{eqnarray}
Also, two parameters of $\phi_0=\phi(r_+,\alpha)$ and $\delta_0=\delta(r_+,\alpha)$ at the outer horizon  will be
determined when matching with an asymptotically flat solution existing  in the far region
\begin{eqnarray}\label{ncoef}
&&m(r)=M-\frac{q^2+q_s^2}{2r}+\cdots,\quad
\delta(r)=\frac{q_s^2}{2r^2}+\cdots,\quad
\phi(r)=\frac{q_s}{r}+\cdots,
\end{eqnarray}
where  $q_s$ represents a primary scalar charge, $M$ is  ADM mass $M$, and $q$ denotes the magnetic charge.

To obatin scalarized black hole solutions, we use the shooting metod to connect the near horizon solutions with the asymptotic solutions. 
Regarding as   explicit scalarized charged black hole solutions,
we present  three numerical black hole solutions with $q=0.5,2,20$ in the $n=0$ fundamental  branch in Fig 6. Each  $N(r)$ traces out  its EEH metric function $f(r)$ for $q=0.5,2,20$ [see (Right) Fig. 1]. Also, we find  different values for the scalar hair depending on $q$ at the horizon. 
Further, it needs  to explore  hundreds of numerical solutions for different  $\alpha$ in the $n=0,1$ branches to perform the stability of  scalarized charged black holes.

\section{Stability for the $n=0,1$ branches }

It is noted that the stability analysis for scalarized charged black holes is an important task
since it determines their viability in representing realistic astrophysical configurations.
The conclusions about the stability of the scalarized charged black holes with respect to  perturbations will be reached by examining
the qualitative behavior of the potential  as well as by obtaining exponentially growing (unstable) modes for $s$-mode scalar  perturbation.
It is known that the $n=0$ fundamental branch is stable agaist radial perturbations, while the exicted ($n=1,2$) branches are unstable in the EMS theory~\cite{Zou:2020zxq}. 

Hence, we prefer to  introduce the radial perturbations around the scalarized black holes as
\begin{eqnarray}
&&ds_{\rm rp}^2=-N(r)e^{-2\delta(r)}(1+\epsilon H_0)dt^2+\frac{dr^2}{N(r)(1+\epsilon H_1)}
+r^2(d\theta^2+\sin^2\theta d\hat{\varphi}^2),\nonumber\\
&&\phi(t,r)=\phi(r)+\epsilon\delta\tilde{\phi}(t,r), \label{p-metric}
\end{eqnarray}
where $N(r)$, $\delta(r)$, and  $\phi(r)$ represent a scalarized
charged black hole background, whereas
$H_0(t,r)$, $H_1(t,r)$, and $\delta\tilde{\phi}(t,r)$
denote three perturbed fields around the scalarized
black hole background. Here, we do not need to introduce a perturbation for the gauge field $A_{\hat{\varphi}}$.
From now on, we focus on  the $l=0$(s-mode)
scalar propagation  by mentioning  that higher angular momentum modes $(l\neq0)$ are neglected. In this
case, other two perturbed fields  become redundant fields.
When looking  for a decoupling process by using  linearized equations, one may find a linearized scalar equation.

Considering the separation of variables
\begin{eqnarray}
\delta\tilde{\phi}(t,r)=\frac{\tilde{\varphi}(r)e^{\Omega t}}{r},
\end{eqnarray}
we obtain the Schr\"odinger-type equation for an $s$-mode scalar perturbation
\begin{eqnarray} \label{sEEH-eq}
\frac{d^2\tilde{\varphi}(r)}{dr_*^2}-\Big[\Omega^2+V_{\rm sEEH}(r,q,\alpha)\Big]\tilde{\varphi}(r)=0,
\end{eqnarray}
with $r_*$ is the tortoise coordinate defined by
\begin{eqnarray}
\frac{dr_*}{dr}=\frac{e^{\delta(r)}}{N(r)}.
\end{eqnarray}
Here, its potential reads to be

\begin{align} \label{sc-poten}
V_{\rm sEEH}(r,q,\alpha)&=
\frac{e^{-2\delta(r) - \alpha \phi^2(r)} \, N(r)}{r^8}
\left[
e^{\alpha \phi^2(r) } r^6
- q^2 r^4
- q^2 r^4 \alpha
+ 2 e^{\alpha \phi^2(r)} q^4 \mu
\right. \nonumber\\
&\quad \left.- e^{\alpha \phi^2(r) } r^6 N(r)
+ 2 q^2 r^4 \alpha^2 \phi^2(r)
- 4 q^2 r^5 \alpha \phi(r) \phi'(r)
+ q^2 r^6 \phi'^2(r)
\right. \nonumber\\
&\quad
- 2 e^{\alpha \phi^2(r)} r^8 \phi'^2(r)
- 2 e^{\alpha \phi^2(r)} q^4 r^2 \mu \phi'^2(r)\Big].
\end{align}

We check that $V_{\rm sEEH}(r,q,\alpha)$ with  $\delta(r)=\phi(r)=0$ and  $N(r)\to f(r)$ reduces to $V_{\rm EES}(r,M,q,\alpha)$ in Eq.(\ref{EEH-P}).
\begin{figure*}[t!]
\centering
\includegraphics[width=0.3\textwidth]{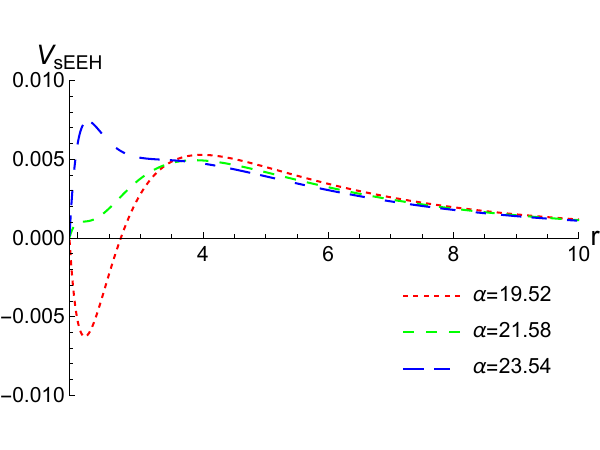}
 \hfill%
\includegraphics[width=0.3\textwidth]{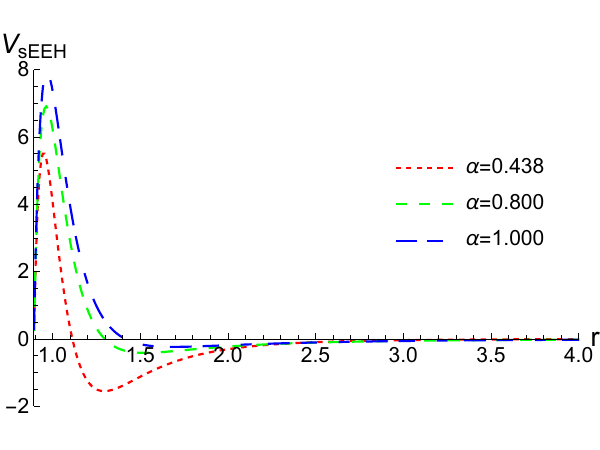}
\hfill%
\includegraphics[width=0.3\textwidth]{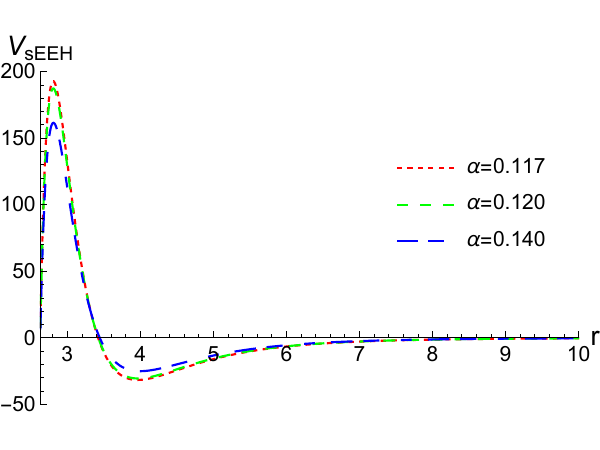}
\caption{ (Left) Three scalar potentials $V_{\rm sEEH}(r,q=0.5,\alpha)$  for $\alpha=19.52,21.58,23.54$ around the $n = 0$ branch.  (Middle)
Three scalar potentials $V_{\rm sEEH}(r,q=2,\alpha)$  for $\alpha=0.438,0.8,1$. (Right) Three scalar potentials $V_{\rm sEEH}(r,q=20,\alpha)$  for $\alpha=0.117,0.12,0.14$. 
Even though they  contain small negative regions in the near horizon, these  turn out to be  stable black holes. }
\end{figure*}
At this stage, we wish to observe the potential $V_{\rm sEEH}(r,q,\alpha)$.
We display three scalar potentials $V_{\rm sEEH}(r,q,\alpha)$  for $q=0.5,2,20$ in Fig. 7 for $l=0(s$-mode) scalar around the $n=0$  branch,  showing  small negative regions near the horizon.
However, this does not imply that the $n=0$ branch is  unstable against the $s$-mode of perturbed scalar because  the sufficient condition for tachyonic instability~\cite{Dotti:2004sh} is given by $\int_{r_+}^\infty dr[e^\delta V_{\rm sEEH}(r,q,\alpha)/N(r)]<0$.
It suggests that the $n=0$ branch  may be  stable against the $s$-mode scalar perturbation.

\begin{figure*}[t!]
   \centering
  \includegraphics[width=0.3\textwidth]{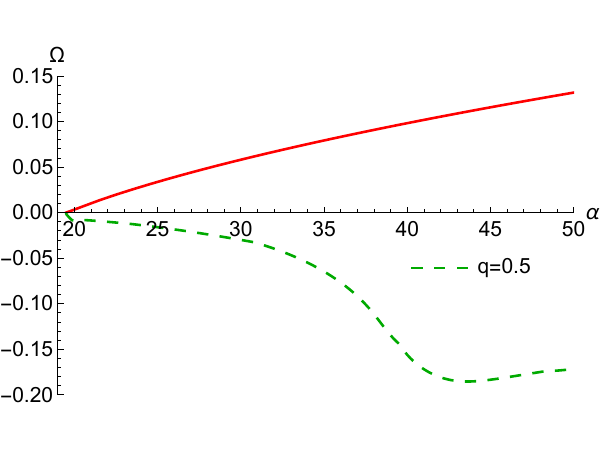}
  \hfill%
\includegraphics[width=0.3\textwidth]{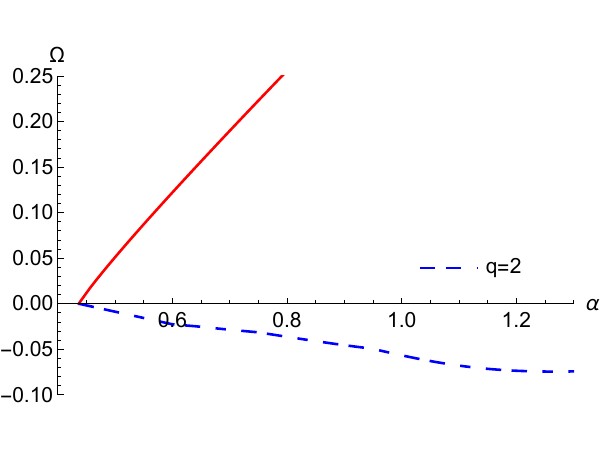}
\hfill%
\includegraphics[width=0.3\textwidth]{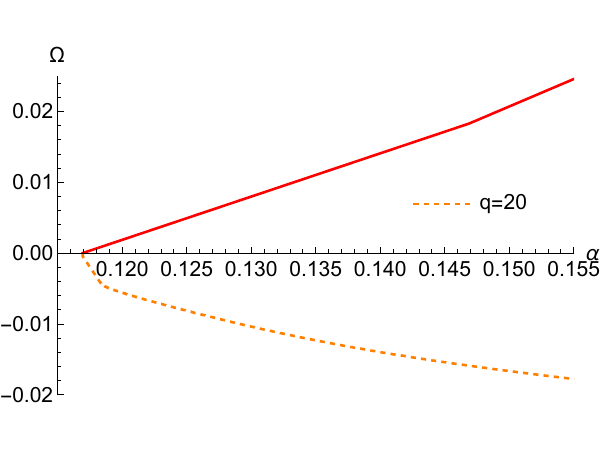}
\caption{Negative $\Omega$ is given as a function of $\alpha$ for the $l = 0$ scalar mode around the $n = 0$ branch, showing stability. Here, we consider three different cases of $q = 0.5$, $2$, and $20$. Three dotted curves start from $\alpha_{n=0} = 19.5102(q=0.5)$, $0.437(q=2)$, and $0.1169(q=20)$. Three red lines represent the unstable EEH black holes [see (Left) Fig. 4].}
\end{figure*}

\begin{figure*}[t!]
   \centering
  \includegraphics[width=0.3\textwidth]{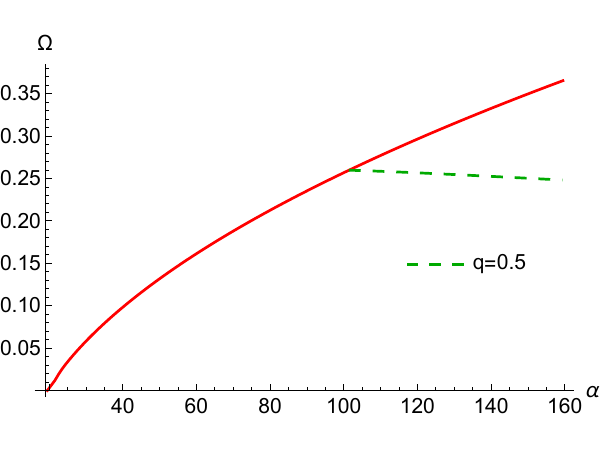}
  \hfill%
\includegraphics[width=0.3\textwidth]{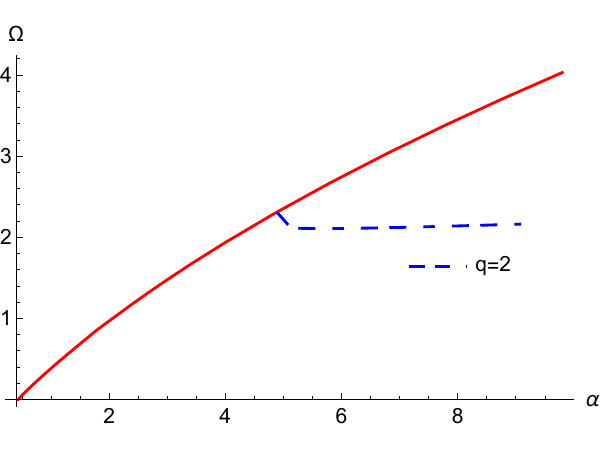}
\hfill%
\includegraphics[width=0.3\textwidth]{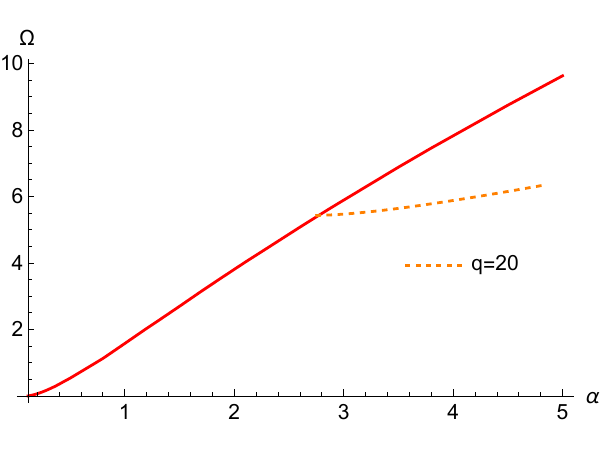}
\caption{Positive  $\Omega$ is given as a function of $\alpha$ for the $l = 0$ scalar mode around the $n = 1$ branch, showing instability. Here, we consider three different cases of $q = 0.5$, $2$, and $20$. Three dotted curves start from the $n=1$ branch points of $\alpha_{n=1} = 101.40(q=0.5)$, $4.888(q=2)$, and $2.741(q=20)$. Three red lines represent the unstable EEH black holes [see (Left) Fig. 4].}
\end{figure*}
We have to solve Eq.(\ref{sEEH-eq}) with two boundary conditions to test the stability of scalarized EEH black holes.
This admits an exponentially growing mode of $e^{\Omega t}$ as an unstable mode.
Actually, we confirm from Fig. 8 that the negative $\Omega$ with three different values of $q = 0.5,2, 20$ implies stability for  the $n = 0$ branch of sEEH black holes.  This shows that the stability of $n=0$ branch is independent of the magnetic charge $q$.  
Three red curves starting at $\alpha= \alpha_{\rm th}= 19.5102(q=0.5),~ 0.437 (2),~ 0.1169(20)$
denote the positive $\Omega$, indicating  the unstable EEH black holes for $\alpha> \alpha_{\rm th}=19.5102(q=0.5),~ 0.437 (2),~ 0.1169(20)$ as was shown in  (Left) Fig. 4.

In addition, we find from Fig. 9 that the $n=1$ branch  with $q=0.5,2,20$ is unstable against radial perturbations.  

\section{Scalarized EEH black holes with both scalar couplings}

Up to now, we considered the case that  scalar field is coupled  to the ordinary Maxwell term $\mathcal{F}$, while $\mathcal{F}^2$ remains uncoupled. From a theoretical viewpoint, it is  natural for the scalar field to couple to both terms. 
Hence,  it is better to consider both scalar couplings to derive scalarized EEH black holes.
First of all, its  linearized scalar equation is given by 
\begin{equation}
\Big[\bar{\nabla}^2+\tilde{\alpha}\Big(\frac{ q^2}{r^4}-2\mu\frac{q^4}{r^8}\Big)\Big]\delta \varphi=0.\label{per6-eq}
\end{equation}
Since $\mu=0.3$-term is a negative term, we expect to have the larger $\tilde{\alpha}_{\rm th}(1,q)$ than $\alpha_{\rm th}(1,q)$ for the single scalar coupling to $\mathcal{F}$.
It is found that  $\tilde{\alpha}_{\rm th}(1,q=0.5) =19.6269(q=0.5)>\alpha_{\rm th}(1,q=0.5)=19.5102$ when the curve crosses the $\alpha$-axis in (Left) Fig. 10. 
\begin{figure*}[t!]
   \centering
  \includegraphics[width=0.4\textwidth]{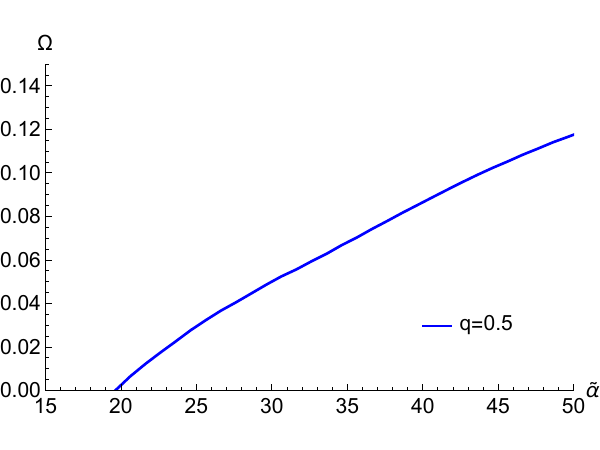}
  \hfill%
  \includegraphics[width=0.4\textwidth]{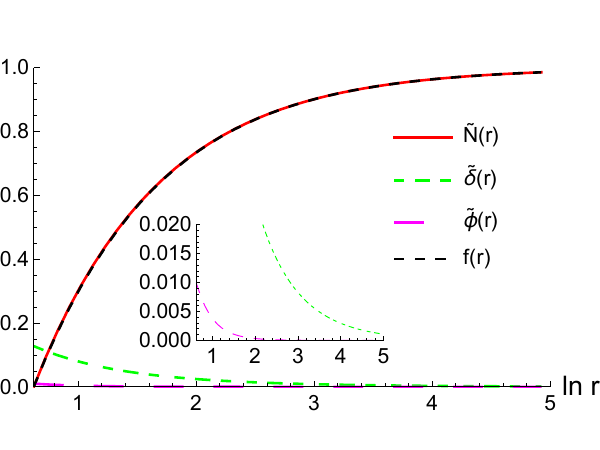}
  \caption{(Left) A curve of $\Omega$ in $e^{\Omega t}$ as a function of $\alpha$ are used to determine the thresholds of tachyonic instability $[\tilde{\alpha}_{\rm th}(1,q)]$ around the EEH black holes for both scalar couplings.
We find that  $\tilde{\alpha}_{\rm th}(1,q=0.5) =19.6269$ when the curve crosses the $\alpha$-axis.
(Right) Graph  of scalarized  black hole  solutions for both scalar couplings. It shows metric functions $\tilde{N}(r)$, $\tilde{\delta}(r)$, and scalar hair $\tilde{\phi}(r)$. 
This graph is found for  $q = 0.5$,  and $\ln r_+ =0.624$ for $\alpha = 25$  in the $n = 0$ branch of $\tilde{\alpha} \geq 19.6269$.}
\end{figure*}
To obtain scalarized black hole solution, we consider   three coupled equations for $\tilde{m}(r),$ $\tilde{\delta}(r),~\tilde{\phi}(r)$ as
\begin{eqnarray}
&&\Big(q^2-\frac{2\mu q^4}{r^4}\Big)e^{-\tilde{\alpha}\tilde{\phi}^2(r)}-2r^2\tilde{m}'(r)+r^3\left(r-2\tilde{m}(r)\right)\tilde{\phi}'^2(r)=0, \label{neom1}\\
&&\tilde{\delta}'(r)+r\tilde{\phi}'^2(r)=0, \label{neom2}\\
&&\tilde{\alpha} \Big(q^2-\frac{2\mu q^4}{r^4}\Big)\frac{\tilde{\phi}(r)e^{-\tilde{\alpha}\tilde{\phi}^2(r)}}{r^2}-2[\tilde{m}(r)+r\tilde{m}'(r)-r]\tilde{\phi}'(r) \nonumber \\
&&\quad\quad-r(r-2\tilde{m}(r))[\tilde{\delta}'(r)\tilde{\phi}'(r)-\tilde{\phi}''(r)]=0. \label{neom3}
\end{eqnarray}
Here, we wish to find the $n=0$ branch of  scalarized  black holes. 
We present  one  numerical black hole solution with  $q=0.5$ for the $n=0$ fundamental  branch in (Right) Fig. 10, which is very similar to scalarized EEH black hole solutions in (Left) Fig. 6.
However, it is challaging to find $q>1$ scalarized black hole solutions.

\section{Discussions }

In this work, we have  explored the spontaneous scalarization of the EEH black hole in the EEHS theory with a scalar coupling function $f(\phi)=e^{-\alpha \phi^2}$ to Maxwell term $\mathcal{F}$.
In this case, there is no restiction on magentic charge $q$ for the choice of  the action parameter $\mu=0.3$.  This means that the overcharge $q>1$ for spontaneous scalarization in the EMS theory
~\cite{Herdeiro:2018wub} disappeared and no-scalar-haired inner horizon theorem~\cite{Cai:2020wrp} can be avoided automatically. 
The computational process is as follows: detecting the tachyonic instability of EEH black holes $\rightarrow$ predicting scalarized EEH black holes (bifurcation points) $\rightarrow$ obtaining the $n = 0,1$ branches of sEEH black holes $\rightarrow$ performing the (in)stability analysis of these branches.

We first point out that the EEH black hole is unstable for $\alpha > \alpha_{\rm th}(q)$ (see Fig. 3), while it remains stable for $\alpha < \alpha_{\rm th}(q)$. The parameter $\alpha_{\rm th}(q)=\alpha_{n=0}(q)$ represents the instability threshold of the EEH black hole and marks the boundary between EEH black holes and the $n=0$ branch of sEEH black holes.
Accordingly, the $n=0$ branch exists for any $\alpha \geq \alpha_{\rm th}(q)$. We further found that the bifurcation point $\alpha_{n=0}(q)$ grows as $q$ decreases, implying that tachyonic instability is harder to be triggered for smaller magnetic charges.  Since all sEEH black hole solutions emerge from spontaneous scalarization, we expect to obtain  infinitely many branches ($n=0,1,2,\cdots$). However, previous works~\cite{Myung:2018vug,Myung:2018jvi} indicate that all existed branches with $n\neq 0$ are unstable under radial perturbations.

Imporatntly, we have shown that when $q = 0.5, 2, 20$, the $n = 0$ branch of sEEH is stable against radial perturbations (see Fig. 8). This may be an issue worth further consideration in future work. Since the $n = 0$ branch of sEEH is stable for $q = 0.5, 2, 20$, and its observational implications may therefore arise~\cite{Stuchlik:2019}. On the other hand, it is worth noting that  a negative potential-induced scalarization of EEH black hole has  led to a single branch of unstable scalarized EEH black holes~\cite{Guo:2025ksj}. 
In addition,  we found from Fig. 9 that the $n=1$ branch  with $q=0.5,2,20$ is unstable against radial perturbations.   Hence, we may deduce that  all existed branches with $n\neq 0$ are unstable under radial perturbations.

Finally, we have discussed the  onset of scalarization for electrically charged black holes, which is similar to that for magnetically charged black holes. Furthermore, we have investigated the onset scalarization of EEH black holes by considering both scalar couplings to  $\mathcal{F}$ and $\mathcal{F}^2$. Its threshold values $\tilde{\alpha}_{\rm th}(1,q=0.5)$ are increased when comparing
 $\alpha_{\rm th}(1,q=0.5)$ for the single scalar coupling to $\mathcal{F}$. Also, its scalarized black hole solutions in the $n=0$ branch are very similar to that for  the single scalar coupling to $\mathcal{F}$. 

 \vspace{1cm}

{\bf Acknowledgments}

 Y.S.M. was supported by the National Research Foundation of Korea (NRF) grant funded by the Korea government(MSIT) (RS-2022-NR069013).  D.C.Z was supported by the National Natural Science Foundation of China (NNSFC) (Grant No.12365009)
 
\end{document}